%% file: main.tex
\documentclass[letterpaper, 10 pt, conference]{ieeeconf} 

\IEEEoverridecommandlockouts 
\input{packages.tex}

\title{\LARGE \bf The Cubli: Modeling and Nonlinear Control \\ Utilizing Unit Complex Numbers}

\author{Fabio Bobrow, Bruno A. Angelico and Paulo S. P. da Silva
\thanks{This work was supported by the S\~ao Paulo Research Foundation (FAPESP) for the grant 2017/22130-4.}
\thanks{The authors are with the Department of Telecommunications and Control Engineer, Escola Polit\'ecnica da USP, S\~ao Paulo, SP, Brazil. The contact author is Fabio Bobrow, e-mail: fbob@usp.br}}

\begin{document}

    \maketitle
    \thispagestyle{empty}
    \pagestyle{empty}

    \begin{abstract}
        This paper covers the modeling and nonlinear control of the Cubli, a cube with three reaction wheels mounted on orthogonal faces that becomes a reaction wheel-based 1D/3D inverted pendulum when positioned in one of its edges (1D) or vertices (3D). Instead of angles, unit complex numbers are used as control states for the 1D configuration. This approach is useful not only to get rid of trigonometric functions, but mainly because it is a specific case of the 3D configuration, that utilizes unit ultra-complex numbers (quaternions) as system states, and therefore facilitates its understanding. The derived nonlinear control law is equivalent to a linear one and is characterized by only three straightforward tuning parameters. Experiment results are presented to validate modeling and control.
    \end{abstract}
    
    \section{INTRODUCTION}

        Inverted pendulum systems have been a popular demonstration of using feedback control to stabilize open-loop unstable systems. Introduced back in 1908 by Stephenson \cite{stephenson_1908_stability}, the first solution to this problem was presented only in 1960 with Roberge \cite{roberge_1960_mechanical} and it is still widely used to test, demonstrate and benchmark new control concepts and theories \cite{xu_2013_kernel}, \cite{yang_2014_neural}, \cite{shi_2015_network}, \cite{wang_2017_smc}. 
        
\begin{figure}[H]
            \centering
            \subfigure[][Cart-pole]{\input{tikz/inverted_pendulum_cart_pole.tex}}
            \subfigure[][Reaction wheel]{\input{tikz/inverted_pendulum_reaction_wheel.tex}} 
            \caption{Inverted pendulum types} 
            \label{fig/inverted_pendulum_types}
        \end{figure}
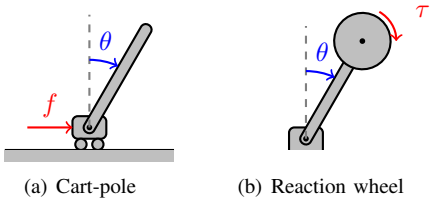

        Differently from cart-pole inverted pendulums, that have a controlled cart with linear motion (Fig. \ref{fig/inverted_pendulum_types}a), reaction wheel pendulums have a controlled rotating wheel that exchanges angular momentum with the pendulum (Fig. \ref{fig/inverted_pendulum_types}b). First introduced in 2001 by Spong \cite{spong_2001_reaction}, it was soon adapted to 3D design variants \cite{sanyal_2004_dynamics,sanyal_2007_rmp}.
        
\begin{figure}
            \centering
            \subfigure[][Edge - 1D]{\includegraphics[width=0.20\textwidth]{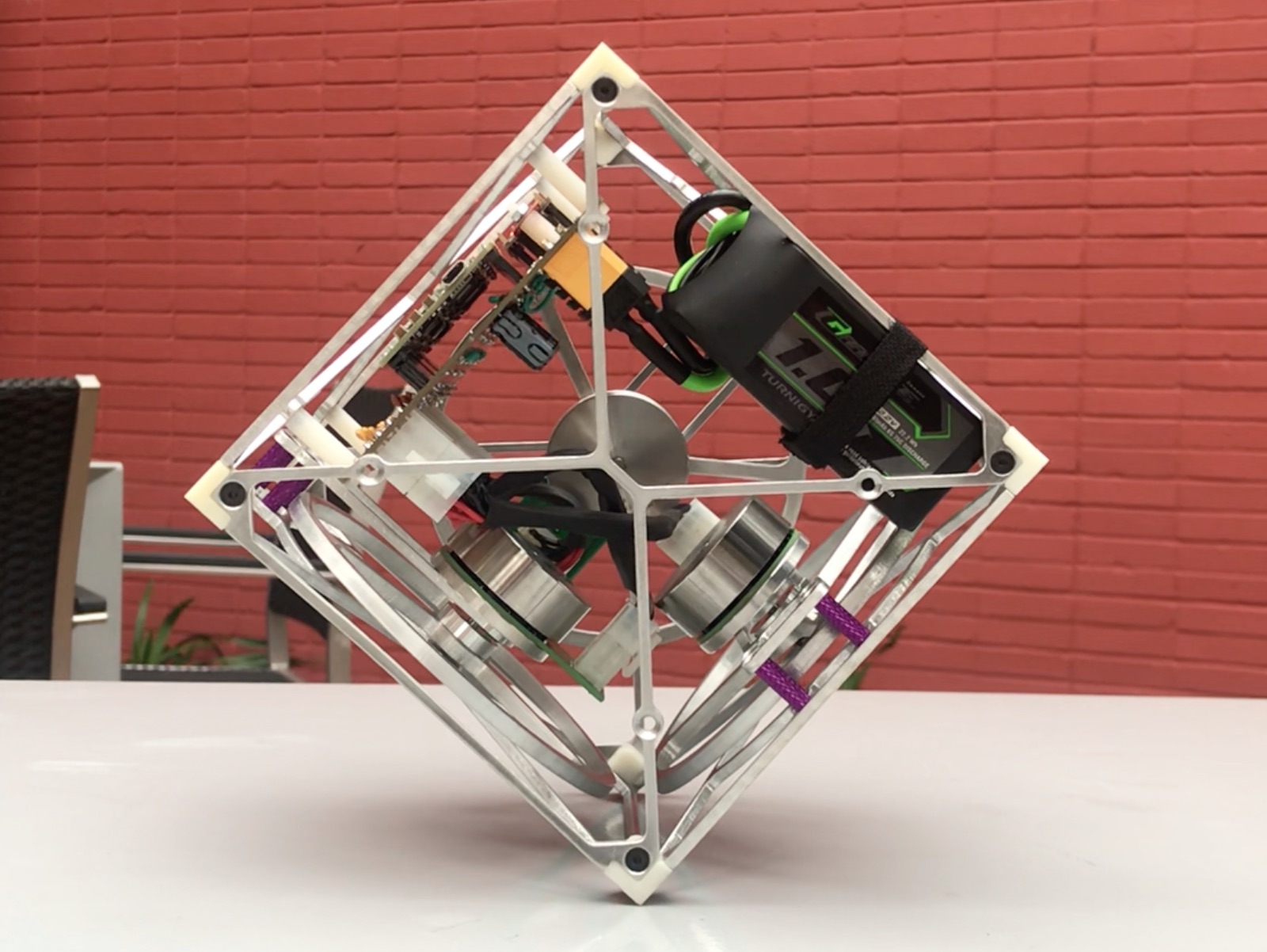}}
            \subfigure[][Vertex - 3D]{\includegraphics[width=0.20\textwidth]{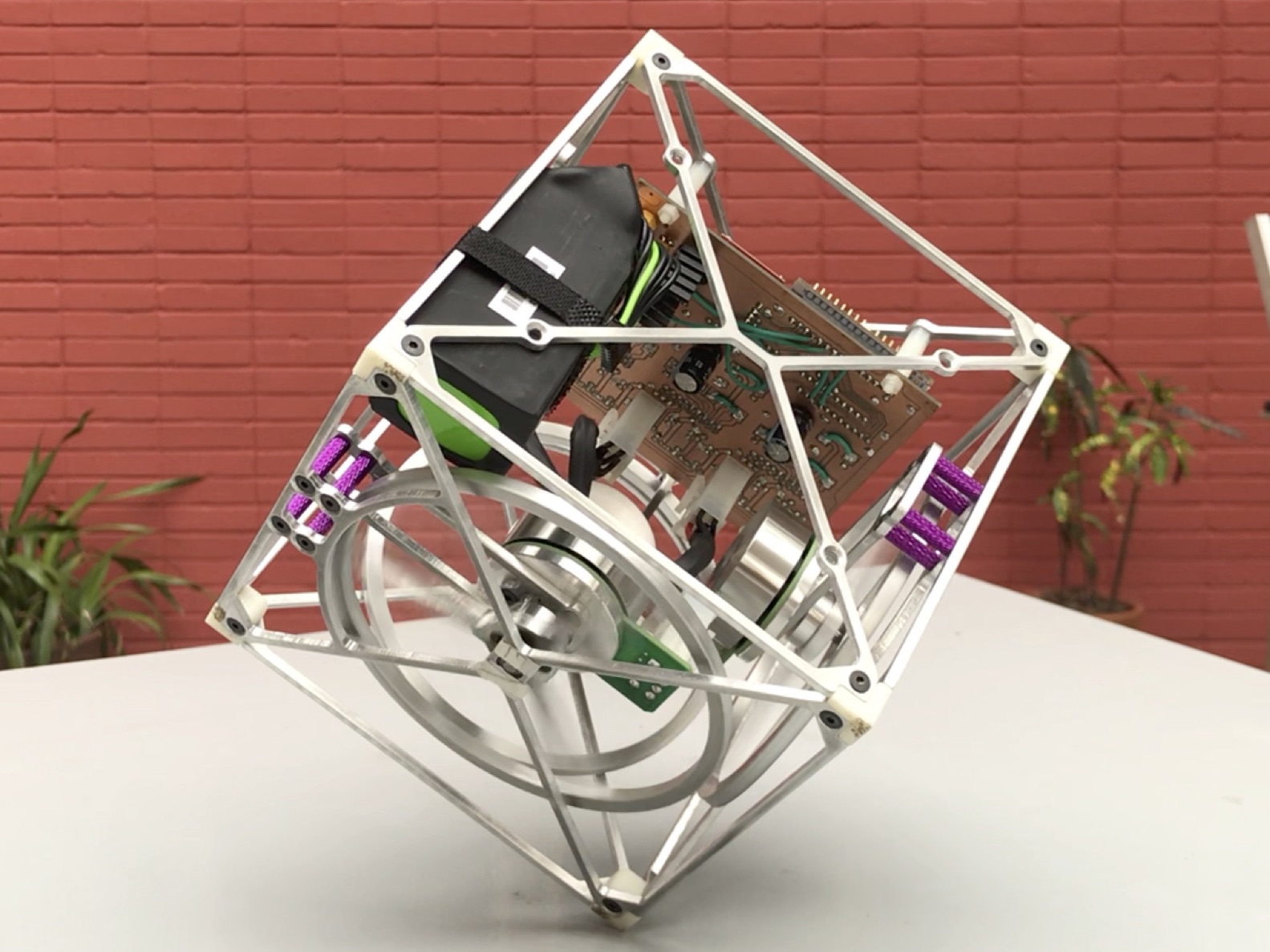}} 
            \caption{Cubli} 
            \label{fig/cubli}
        \end{figure}

        Perhaps, the most notable of them is the Cubli. Originally developed and baptized in 2012 by Gajamohan \cite{gajamohan_2012_cubli,gajamohan_2013_cubli_1} and Muehlebach \cite{muehlebach_2013_cubli,muehlebach_2017_cubli} from the Institute for Dynamic Systems and Control of Zurich Federal Institute of Technology (ETH Zurich), the Cubli is a device that consist of a cube with three reaction wheels mounted on orthogonal faces. By positioning the Cubli on its edge, it becomes a reaction wheel-based 1D inverted pendulum (Fig. \ref{fig/cubli}a), while if it is positioned on its vertex, it becomes a reaction wheel-based 3D inverted pendulum (Fig. \ref{fig/cubli}b).
        
        The purpose of this paper is to model the 1D configuration and then design and implement a nonlinear controller for it. Although this is widely available in the literature \cite{fantoni_2001_stabilization}, \cite{jepsen_2009_development}, the approach in this paper utilizes unit complex numbers as control states instead of angles. 
        
\begin{figure}[H]
            \centering
            \input{tikz/complex_number.tex}
            \caption{Unit complex number}
            \label{fig/complex_number}
        \end{figure}
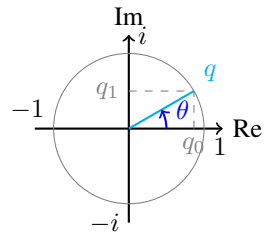

        Unit complex numbers  (Fig. \ref{fig/complex_number}), also called unit circle $\mathbb{S}^1$, circle group $\mathbb{T}^1$ or special orthogonal group $SO(2)$, are the multiplicative group of all complex numbers with absolute value $1$. They form a commutative compact Lie Group, with planar 2D rotation employed as a group operator, that is well-known and has been employed in several fields, for instance, to describe the synchronization behavior of Kuramoto oscillators \cite{chopra_2005_synchronization}. Due to its Lie Group structure, they have been exploited in the literature to yield global/almost global results, often through simple Lyapunov-based analysis \cite{bosso_2019_global}.
        
        Even though this approach discards the use of trigonometric functions, the main goal here is to facilitate the understanding of the 3D configuration, which is a more generic case that utilizes unit ultra-complex numbers (quaternions) as system states. There, the unit circle $\mathbb{S}^1$ in $\mathbb{R}^2$ gets replaced by the unit 3-sphere $\mathbb{S}^3$ in $\mathbb{R}^4$, which cannot be easily visualized in our 3D world. Moreover, they are a non-commutative compact Lie Group.
    
    \section{SYSTEM MODELING}
    
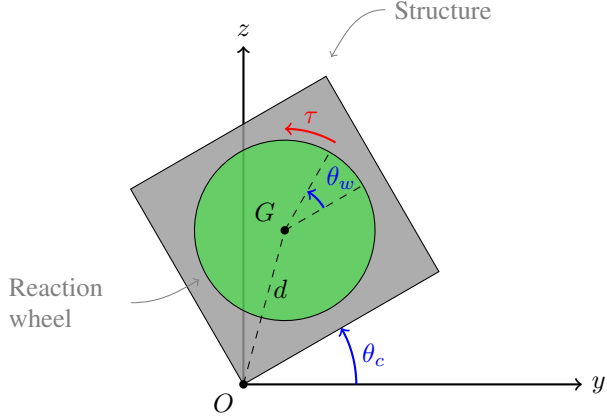
\begin{figure}[H]
                \centerline{\input{tikz/cubli_diagram.tex}}
                \caption{Cubli schematic diagram} \label{fig/cubli_diagram}
            \end{figure}

        The Cubli balancing on its edge is composed of two rigid bodies: a structure and a reaction wheel (Fig. \ref{fig/cubli_diagram}). The structure rotates freely around the pivot point $O$ (articulation edge), while the reaction wheel, besides rotating together with the structure, also rotates around its center of mass $G$ (axial axis). 
            
        \subsection{Notations}
            
            Let $\theta_c$ and $\omega_c$ denote the structure angular displacement and velocity. Let $\theta_w$ and $\omega_w$ denote the reaction wheel relative (measured with respect to the structure) angular displacement and velocity.
        
            Let $l$ denote the structure side length, $m_s$ denote the structure mass and $I_{s_{G}}$ denote the structure moment of inertia around its center of mass $G$. Let $m_w$ denote the reaction wheel mass and $I_{w_{G}}$ denote the reaction wheel moment of inertial around its center of mass $G$. These parameters were obtained from the CAD version of the Cubli and are given in Tab. \ref{tab:cubli_parameters}.
            
            \begin{table}[H]
                \caption{The Cubli parameters}
                \label{tab:cubli_parameters}
                \centering
                \begin{tabular}{|cc|}
                    \hline
                    Parameter & Value \\
                    \hline
                    $l$ & $ 0.15$m \\
                    $m_s$ & $0.70$ kg \\
                    $m_w$ & $0.15$ kg \\
                    \hline
                \end{tabular}
                \hspace{0.2cm}
                \begin{tabular}{|cc|}
                    \hline
                    Parameter & Value \\
                    \hline
                    $I_{s_G}$ & $3.75\times10^{-3}$ kg.m$^2$ \\
                    $I_{w_G}$ & $1.25\times10^{-4}$ kg.m$^2$ \\
                    \hline
                \end{tabular}
            \end{table}
            
            The constant $d = \frac{l\sqrt{2}}{2}$ is the distance between pivot point $O$ and structure/reaction wheel center of mass $G$, $I_{s_O} = I_{s_G} + m_s d^2$ and $I_{w_O} = I_{w_G} + m_w d^2$ are the structure and reaction wheel moment of inertia around pivot point $O$, whereas $I_{c_O} = I_{s_O} + I_{w_O}$ represents the Cubli total moment of inertia around pivot point $O$, $m_c = m_s + m_w$ denotes the Cubli total mass and $g$ denote the acceleration of gravity.
            
        \subsection{Kinectic energy}
        
            The structure kinetic energy $T_s$ is given by:
            \begin{align}
                T_s &= \frac{1}{2} I_{s_G} \dot{\theta}_c^2 + \frac{1}{2} m_s {\left( d\dot{\theta}_c \right)}^2 \nonumber \\
                T_s &= \frac{1}{2} \left( I_{s_G} \dot{\theta}_c^2 + m_s d^2 \right) \dot{\theta}_c^2 \nonumber \\
                T_s &= \frac{1}{2} I_{s_O} \dot{\theta}_c^2
                \label{eqn/kinetic_energy_structure}
            \end{align}
        
            The reaction wheel kinetic energy $T_w$ is given by:
            \begin{align}
                T_w &= \frac{1}{2} I_{w_G} {\left( \dot{\theta}_c^2 + \dot{\theta}_w^2 \right)}^2 + \frac{1}{2} m_w {\left( d\dot{\theta}_c \right)}^2 \nonumber \\
                T_w &= \frac{1}{2} I_{w_G} {\left( \dot{\theta}_c^2 + \dot{\theta}_w^2 \right)}^2 + \frac{1}{2} m_w d^2 \dot{\theta}_c^2 \nonumber \\
                T_w &= \frac{1}{2} I_{w_G} {\left( \dot{\theta}_c^2 + \dot{\theta}_w^2 \right)}^2 + \frac{1}{2} \left( I_{w_O} - I_{w_G} \right) \dot{\theta}_c^2
                \label{eqn/kinetic_energy_reaction_wheel}
            \end{align}
            
            Thus, the Cubli total kinetic energy $T$ is the sum of Eqn. (\ref{eqn/kinetic_energy_structure}) and (\ref{eqn/kinetic_energy_reaction_wheel}):
            \begin{align}
                T &= \frac{1}{2} I_{s_O} \dot{\theta}_c^2 + \frac{1}{2} I_{w_G} {\left( \dot{\theta}_c^2 + \dot{\theta}_w^2 \right)}^2 + \frac{1}{2} \left( I_{w_O} - I_{w_G} \right) \dot{\theta}_c^2 \nonumber \\
                T &= \frac{1}{2} \left( I_{s_O} + I_{w_O} - I_{w_G} \right) \dot{\theta}_c^2 + \frac{1}{2} I_{w_G} {\left( \dot{\theta}_c^2 + \dot{\theta}_w^2 \right)}^2 \nonumber \\
                T &= \frac{1}{2} \left( I_{c_O} - I_{w_G} \right) \dot{\theta}_c^2 + \frac{1}{2} I_{w_G} {\left( \dot{\theta}_c^2 + \dot{\theta}_w^2 \right)}^2 \nonumber \\
                T &= \frac{1}{2} \bar{I}_{c_O} \dot{\theta}_c^2 + \frac{1}{2} I_{w_G} {\left( \dot{\theta}_c^2 + \dot{\theta}_w^2 \right)}^2 
                \label{eqn/kinetic_energy}
            \end{align}
            
            \noindent
            where $\bar{I}_{c_O} = I_{c_O} - I_{w_G}$ is the Cubli total moment of inertial around pivot point $O$ without the reaction wheel moment of inertial around its center of mass $G$.
            
        \subsection{Potential energy}
        
            The structure potential energy $V_s$ is given by:
            \begin{equation}
                V_s = m_s g d \sin \left( \theta_c + \frac{\pi}{4} \right)
                \label{eqn/potential_energy_structure}
            \end{equation}
        
            The reaction wheel potential energy $V_w$ is given by:
            \begin{equation}
                V_w = m_w g d \sin \left( \theta_c + \frac{\pi}{4} \right)
                \label{eqn/potential_energy_reaction_wheel}
            \end{equation}
            
            Thus, the Cubli total potential energy $V$ is the sum of Eqn. (\ref{eqn/potential_energy_structure}) and (\ref{eqn/potential_energy_reaction_wheel}):
            \begin{align}
                V_s &= m_s g d \sin \left( \theta_c + \frac{\pi}{4} \right) + m_w g d \sin \left( \theta_c + \frac{\pi}{4} \right) \nonumber \\
                V_s &= \left( m_s + m_w \right) g d \sin \left( \theta_c + \frac{\pi}{4} \right) \nonumber \\
                V_s &= m_c g d \sin \left( \theta_c + \frac{\pi}{4} \right)
                \label{eqn/potential_energy}
            \end{align}
            
        \subsection{Equations of motion}
        
            Once the kinetic and potential energy have been defined, the equations of motion can be derived utilizing Lagrange equations. Let $\tau$ denote the input torque of the motor and $\tau_f(\dot{\theta}_w)$ denote the non-linear friction torque of the motor (to be detailed further).
            
            For the generalized coordinate $\theta_c$:
            \begin{align}
                \frac{d}{dt} \left( \frac{\partial T}{\partial \dot{\theta}_c} \right) - \frac{\partial T}{\partial \theta_c} + \frac{\partial V}{\partial \theta_c} &= Q_{\theta_c} \nonumber \\
                \bar{I}_{c_O} \ddot{\theta}_c^2 + I_{w_G} \left( \ddot{\theta}_c^2 + \ddot{\theta}_w^2 \right) + m_c g d \cos \theta_c &= 0 
                \label{eqn/lagrange_equation_structure}
            \end{align}
            
            For the generalized coordinate $\theta_w$:
            \begin{align}
                \frac{d}{dt} \left( \frac{\partial T}{\partial \dot{\theta}_w} \right) - \frac{\partial T}{\partial \theta_w} + \frac{\partial V}{\partial \theta_w} &= Q_{\theta_w} \nonumber \\
                I_{w_G} \left( \ddot{\theta}_c^2 + \ddot{\theta}_w^2 \right) &=  -\tau_f ( \dot{\theta}_w ) + \tau
                \label{eqn/lagrange_equation_reaction_wheel}
            \end{align}
            
            The Cubli equations of motion are composed of Eqn. (\ref{eqn/lagrange_equation_structure}) and (\ref{eqn/lagrange_equation_reaction_wheel}). They can be written together in matrix notation with the time derivative terms in evidence:
            \begin{equation}
                \begin{bmatrix}
                    \bar{I}_{c_O} & I_{w_G} \\
                    0 & I_{w_G}
                \end{bmatrix}
                \begin{bmatrix}
                    \ddot{\theta}_c^2 \\
                    \ddot{\theta}_c^2 + \ddot{\theta}_w^2
                \end{bmatrix}
                =
                \begin{bmatrix}
                    - m_c g d \cos \theta_c  \\
                    - \tau_f ( \dot{\theta}_w ) + \tau  
                \end{bmatrix}
            \end{equation}
            
            Isolating the time derivative terms:
            \begin{align}
                \begin{bmatrix}
                    \ddot{\theta}_c^2 \\
                    \ddot{\theta}_c^2 + \ddot{\theta}_w^2
                \end{bmatrix}
                &=
                {
                \begin{bmatrix}
                    \bar{I}_{c_O} & I_{w_G} \\
                    0 & I_{w_G}
                \end{bmatrix}
                }^{-1}
                \begin{bmatrix}
                    - m_c g d \cos \theta_c  \\
                    - \tau_f ( \dot{\theta}_w ) + \tau 
                \end{bmatrix} \nonumber \\
                \begin{bmatrix}
                    \ddot{\theta}_c^2 \\
                    \ddot{\theta}_c^2 + \ddot{\theta}_w^2
                \end{bmatrix}
                &=
                \begin{bmatrix}
                    \frac{1}{\bar{I}_{c_O}} & - \frac{1}{\bar{I}_{c_O}} \\
                    0 & \frac{1}{I_{w_G}}
                \end{bmatrix}
                \begin{bmatrix}
                    - m_c g d \cos \theta_c  \\
                    - \tau_f ( \dot{\theta}_w ) + \tau
                \end{bmatrix} \nonumber \\
                \begin{bmatrix}
                    \ddot{\theta}_c^2 \\
                    \ddot{\theta}_c^2 + \ddot{\theta}_w^2
                \end{bmatrix}
                &=
                \begin{bmatrix}
                    \frac{1}{\bar{I}_{c_O}} \left( - m_c g d \cos \theta_c + \tau_f ( \dot{\theta}_w ) - \tau \right) \\
                    \frac{1}{I_{w_G}} \left( -\tau_f ( \dot{\theta}_w ) + \tau \right)
                \end{bmatrix}
            \end{align}
        
            Because the Cubli total moment of inertia is significantly larger than the reaction wheel moment of inertia ($\bar{I}_{c_O} \gg I_{w_G}$), the reaction wheel angular acceleration will be significantly larger than the structure angular acceleration ($\ddot{\theta}_w \gg \ddot{\theta}_c$). Thus, the full equations of motion of the system are given by:
            \begin{equation}
                \left\{
                \begin{array}{l}
                    \dot{\theta}_c = \omega_c \\
                    \dot{\theta}_w = \omega_w \\
                    \dot{\omega}_c = \frac{1}{\bar{I}_{c_O}} \left( - m_c g d \cos \theta_c + \tau_f ( \omega_w ) - \tau \right) \\
                    \dot{\omega}_w = \frac{1}{I_{w_G}} \left( -\tau_f ( \omega_w ) + \tau \right)
                \end{array}
                \right.
                \label{eqn:dynamic_equations}
            \end{equation} 
            
\begin{figure}[H]
                \centering
                \input{tikz/plant_cubli_dynamics.tex}
                \caption{Cubli dynamics}
                \label{fig:plant_cubli_dynamics}
            \end{figure}
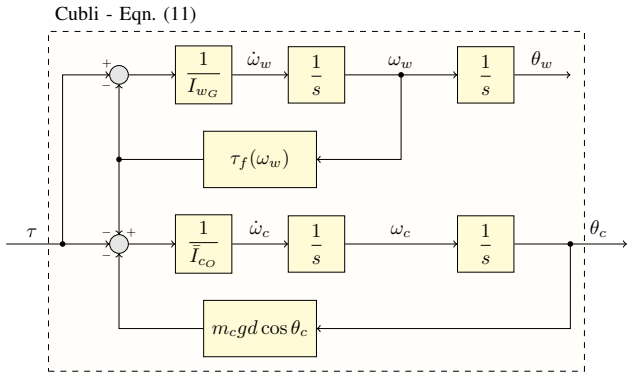

            The system can be represented as a block diagram (Fig. \ref{fig:plant_cubli_dynamics}), where it is easier to interpret the gravity torque and motor friction terms.
            
            Note that the only thing coupling the Cubli and the reaction wheel dynamics is the motor friction.
            
    \section{UNIT COMPLEX NUMBERS}
        
        Instead of using the angle $\theta_c$ to describe the Cubli orientation, a unit complex number $q$ will be considered.
        
        \subsection{Complex number notation}
        
        	A complex number $q$ is a set of two parameters, a real part $q_0$ and an imaginary part $q_1$:
        	\begin{equation}
        		q = q_0 + q_1i
        	\end{equation}
        	
        	\noindent
        	where:
            \begin{equation}
                i^2 = -1
        	    \label{eqn/complex_number_fundamentals}
            \end{equation}
            
            A complex number can also be represented as a two dimension column vector:
            \begin{equation}
            	q =
            	\begin{bmatrix}
            		q_0 \\
                	q_1
            	\end{bmatrix}
            \end{equation}
            
            The conjugate of a complex number is defined as:
            \begin{equation}
            	\bar{q} =
            	\begin{bmatrix}
            		q_0 \\
                	-q_1
            	\end{bmatrix},
            \end{equation}
            
            \noindent
            and its norm (a nonnegative real value) as:
            \begin{equation}
            	|q| = \sqrt{q^T q} = \sqrt{q_0^2+q_1^2}
            \end{equation}
        
        \subsection{Complex number product}
            
            From the rule given in Eqn. (\ref{eqn/complex_number_fundamentals}), the product of two complex numbers $q$ and $r$ (represented by the $\circ$ operator) can be derived:
            \begin{equation}
                q \circ r =
                \begin{bmatrix}
                    q_0 r_0 - q_1 r_1 \\
                    q_0 r_1 + q_1 r_0 
                \end{bmatrix}
                \label{eqn/complex_number_product_1}
            \end{equation}
            
            Since Eqn. (\ref{eqn/complex_number_product_1}) is linear in $r$, it can also be written in matrix-vector product form:
            \begin{equation}
                q \circ r =
                \underbrace{
                \begin{bmatrix}
                    q_0 & -q_1 \\
                    q_1 & q_0
                \end{bmatrix}
                }_{R(q)}
                \begin{bmatrix}
                    r_0 \\
                    r_1 
                \end{bmatrix}
                \label{eqn/complex_number_product_2}
            \end{equation}
            
            \noindent
            where:
            \begin{equation}
                R(q) = 
                \begin{bmatrix}
                    \vert & \vert \\
                    q   & G(q)^T   \\
                    \vert & \vert
                \end{bmatrix}
            \end{equation}
           
            \noindent
            and:
            \begin{equation}
                G(q) = 
                \begin{bmatrix}
                    -q_1 & q_0
                \end{bmatrix}
            \end{equation}
            
            From Eqn. (\ref{eqn/complex_number_product_2}), it turns out that:
            \begin{equation}
                q \circ \bar{q} = \bar{q} \circ q =
                \begin{bmatrix}
                    |q|^2 \\
                    0
                \end{bmatrix},
                \label{eqn/complex_number_property_1}
            \end{equation}
            
            \noindent
            and if the complex number has unitary norm ($|q|=1$):
            \begin{equation}
                q \circ \bar{q} = \bar{q} \circ q =
                \begin{bmatrix}
                    1 \\
                    0
                \end{bmatrix}
                \label{eqn/complex_number_property_2}
            \end{equation}
            
        \subsection{Unit complex number}
        
            Let $q$ be a unit complex number, that is, constrained to have unitary norm ($|q|=1$):
            \begin{equation}
                q^T q = 1
                \label{eqn/rotation_complex_number_constrain}
            \end{equation}
            
            Its real and imaginary parts will be given solely by the angle $\theta$ it makes with the real axis (Fig. \ref{fig/complex_number}):
            \begin{equation}
                q
                =
                \begin{bmatrix}
                    \cos \theta \\
                    \sin \theta
                \end{bmatrix}
                \label{eqn/rotation_complex_number}
            \end{equation}
            
            In other words, a unit complex number $q$ is a redundant way of describing a rotational angle $\theta$.
            
            Differentiating Eqn. (\ref{eqn/rotation_complex_number}), yields:
            \begin{align}
                \dot{q} &= \frac{d}{dt}
                \begin{bmatrix}
                    \cos\theta \\ 
                    \sin\theta
                \end{bmatrix} \nonumber \\
                \dot{q} &= 
                \begin{bmatrix}
                    -\dot{\theta} \sin\theta \\ 
                    \dot\theta \cos\theta
                \end{bmatrix} \nonumber \\
                \dot{q} &= 
                \underbrace{
                \begin{bmatrix}
                    -q_1 \\ 
                    q_0
                \end{bmatrix}
                }_{G(q)^T}
                \omega
                \label{eqn/rotation_kinematic_equation}
            \end{align}
            
            This is the rotation kinematic equation utilizing unit complex numbers.
            
            Left multiplying Eqn. (\ref{eqn/rotation_kinematic_equation}) by $G(q)$, yields:
            \begin{align}
                \begin{bmatrix}
                    -q_1 & q_0
                \end{bmatrix}
                \begin{bmatrix}
                    \dot{q}_0 \\ 
                    \dot{q}_1
                \end{bmatrix}
                &= 
                \begin{bmatrix}
                    -q_1 & q_0
                \end{bmatrix}
                \begin{bmatrix}
                    -q_1 \\ 
                    q_0
                \end{bmatrix}
                \omega \nonumber \\
                \begin{bmatrix}
                    -q_1 & q_0
                \end{bmatrix}
                \begin{bmatrix}
                    \dot{q}_0 \\ 
                    \dot{q}_1
                \end{bmatrix}
                &= 
                \cancel{\left( q_1^2 + q_0^2 \right)}
                \omega \nonumber \\
                \begin{bmatrix}
                    -q_1 & q_0
                \end{bmatrix}
                \begin{bmatrix}
                    \dot{q}_0 \\ 
                    \dot{q}_1
                \end{bmatrix}
                &= 
                \omega
                \label{eqn/rotation_kinematic_equation_2}
            \end{align}
            
            Differentiating the constrain Eqn. (\ref{eqn/rotation_complex_number_constrain}), yields:
            \begin{align}
                \frac{d}{dt} \left( q^T q \right) &= \frac{d}{dt} \left( 1 \right) \nonumber \\
                \dot{q}^T q + q^T \dot{q} &= 0 \nonumber \\
                \cancel{2} q^T \dot{q} &= 0 \nonumber \\
                \begin{bmatrix}
                    q_0 & q_1
                \end{bmatrix}
                \begin{bmatrix}
                    \dot{q}_0 \\ 
                    \dot{q}_1
                \end{bmatrix}
                &= 0 
                \label{eqn/rotation_complex_number_constrain_2}
            \end{align}
            
            Joining together Eqn. (\ref{eqn/rotation_kinematic_equation_2}) and (\ref{eqn/rotation_complex_number_constrain_2}):
            \begin{equation}
                \begin{bmatrix}
                    q_0 & q_1 \\
                    -q_1 & q_0
                \end{bmatrix}
                \begin{bmatrix}
                    \dot{q}_0 \\ 
                    \dot{q}_1
                \end{bmatrix}
                =
                \begin{bmatrix}
                    0 \\ 
                    \omega
                \end{bmatrix}
                \label{eqn/rotation_kinematic_equation_and_constrain}
            \end{equation}
            
            Comparing Eqn. (\ref{eqn/rotation_kinematic_equation_and_constrain}) with (\ref{eqn/complex_number_product_2}), it can be seen that:
            \begin{equation}
                \begin{bmatrix}
                    0 \\ 
                    \omega
                \end{bmatrix}
                = \bar{q} \circ \dot{q}
                = \dot{q} \circ \bar{q}
            \end{equation}
            
            Or also:
            \begin{equation}
                \begin{bmatrix}
                    0 \\ 
                    \omega
                \end{bmatrix}
                = - \dot{\bar{q}} \circ q
                = - q \circ \dot{\bar{q}}
                \label{eqn/inverse_rotation_kinematic_equation}
            \end{equation}
        
        \subsection{Equations of motion}
        
            The Cubli equations of motion from Eqn. (\ref{eqn:dynamic_equations}) can be rewritten in terms of unit complex number $q$:
            \begin{equation}
                \left\{
                \begin{array}{l}
                    \dot{q} = G(q)^T \omega_c \\
                    \dot{\theta}_w = \omega_w \\
                    \dot{\omega}_c = \frac{1}{\bar{I}_{c_O}} \left( - m_c g d \Gamma q + \tau_f ( \omega_w ) - \tau \right) \\
                    \dot{\omega}_w = \frac{1}{I_{w_G}} \left( -\tau_f ( \omega_w ) + \tau \right)
                \end{array}
                \right.
                \label{eqn:dynamic_equations_complex_numbers}
            \end{equation} 
            
            \noindent
            where:
            \begin{equation}
                \Gamma = 
                \begin{bmatrix}
                    1 & 0
                \end{bmatrix}
            \end{equation}
            
\begin{figure}[H]
                \centering
                \input{tikz/plant_cubli_dynamics_complex_numbers.tex}
                \caption{Cubli dynamics (with unit complex numbers)}
                \label{fig:plant_cubli_dynamics_complex_numbers}
            \end{figure}
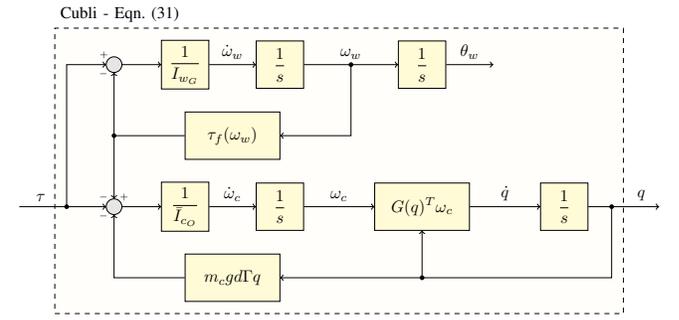

            Despite having one more equation now (since $q$ is a two-dimensional vector), there is no longer any trigonometric function. The block diagram (Fig. \ref{fig:plant_cubli_dynamics_complex_numbers}) is also quite similar.
            
        \subsection{Linearized dynamics}
            
            When the Cubli is at rest $\omega_c = \theta_w  = \omega_w = 0$, perfectly balanced in its unstable equilibrium position $q = q_u = {\begin{bmatrix} \frac{\sqrt{2}}{2} & \frac{\sqrt{2}}{2} \end{bmatrix}}^T$, the linearized dynamics are:
            \begin{equation}
                \resizebox{0.49\textwidth}{!}{$
                \begin{bmatrix}
                    \dot{q} \\
                    \dot{\theta}_w \\
                    \dot{\omega}_c \\
                    \dot{\omega}_w
                \end{bmatrix}
                =
                \left[
                \begin{array}{c;{1pt/2pt}c;{1pt/2pt}c;{1pt/2pt}c}
                    0_{2\times2} & 0_{2\times1} & G^T(q_u) & 0_{2\times1} \\
                    \hdashline[1pt/2pt]
                    0_{1\times2} & 0 & 0 & 1 \\
                    \hdashline[1pt/2pt]
                    -\frac{m_c g d}{\bar{I}_{c_O}} \Gamma & 0 & 0 & \frac{b_w}{\bar{I}_{c_O}} \\
                    \hdashline[1pt/2pt]
                    0_{1\times2} & 0 & 0 & -\frac{b_w}{I_{w_G}}
                \end{array}
                \right]
                \begin{bmatrix}
                    q \\
                    \theta_w \\
                    \omega_c \\
                    \omega_w
                \end{bmatrix}
                +
                \begin{bmatrix}
                    0_{2\times1} \\
                    0 \\
                    -\frac{1}{\bar{I}_{c_O}} \\
                    \frac{1}{I_{w_G}}
                \end{bmatrix}
                \vec{\tau}
                $}
            \end{equation}
            
            Its characteristic polynomial is given by:
            \begin{equation}
                \underbrace{s}_{\begin{array}{c}\text{u. c. n.} \\ \text{redu.}\end{array}} \underbrace{s \left( s + \omega_1 \right)}_{\begin{array}{c}\text{r. wheel} \\ \text{dynamics}\end{array}} \underbrace{\left( s^2 - \omega_0^2 \right)}_{\begin{array}{c}\text{cubli} \\ \text{dynamics}\end{array}} = 0
            \end{equation}
            
            \noindent
            where $\omega_0$ is the natural frequency of the Cubli dynamics, whereas $\omega_1$ is the natural frequency of the reaction wheel dynamics, given by:
            
            \begin{equation}
                \omega_0 = \sqrt{\frac{m_c g d \frac{\sqrt{2}}{2}}{\bar{I}_{c_O}}}
                , \qquad
                \omega_1 = \frac{b_w}{I_{w_G}}
            \end{equation}
            
            The Cubli is an unstable system due to its poles being located at $\pm\omega_0$, while the reaction wheel is marginally stable due to its poles being located at $0$ and $-\omega_1$. Moreover, there is also an extra pole at $0$, which is inherited from the kinematic equation, since a unit complex number is a redundant way to describe an angle.
            
            % \begin{figure}[H]
            %     \centering
            %     \input{tikz/open_loop_poles}
            %     \caption{Open loop poles} 
            %     \label{fig/open_loop_poles}
            % \end{figure}
            
            The controllability matrix has $\text{rank}(\mathcal{C})=4$, while the system has dimension $n=5$. However, even with $\text{rank}(\mathcal{C})\neq n$, the system is full controllable since one of the system states is redundant due to its unit complex number representation. In other words, although unit complex numbers are being utilized (which includes an extra redundant state), the system still have 2 d.o.f. and thus its ``physical'' dimension remains $n=4$.

    \section{ATTITUDE CONTROLLER}
    
        Initially, we will focus only on the Cubli dynamics, without concerning about controlling the reaction wheel.
            
        \subsection{Friction torque compensation}
    
            The friction torque $\tau_f(\omega_w)$ occurs in the opposite direction of the reaction wheel angular velocity $\omega_w$, and it corresponds to Coulomb (static) and viscous (dynamic) friction of the motor. However, because the reaction wheel is hollow, there is also a significant aerodynamic drag. Given that, the friction torque can be approximated with:
            \begin{equation}
                \tau_f(\omega_w) = \text{sign}(\omega_w) \left[ \tau_c + b_w |\omega_w| + c_d |\omega_w|^2 \right]
            \end{equation}
            
            \noindent
            where $\tau_c$ is the Coulomb friction, $b_w$ is the viscous friction coefficient and $c_d$ is the aerodynamic drag coefficient.
            
\begin{figure}[H]
                %\centerline{\includegraphics[width=0.5\textwidth]{fig/friction_torque_wheel_angular_velocity.pdf}}
                \centerline{\includegraphics[width=0.5\textwidth]{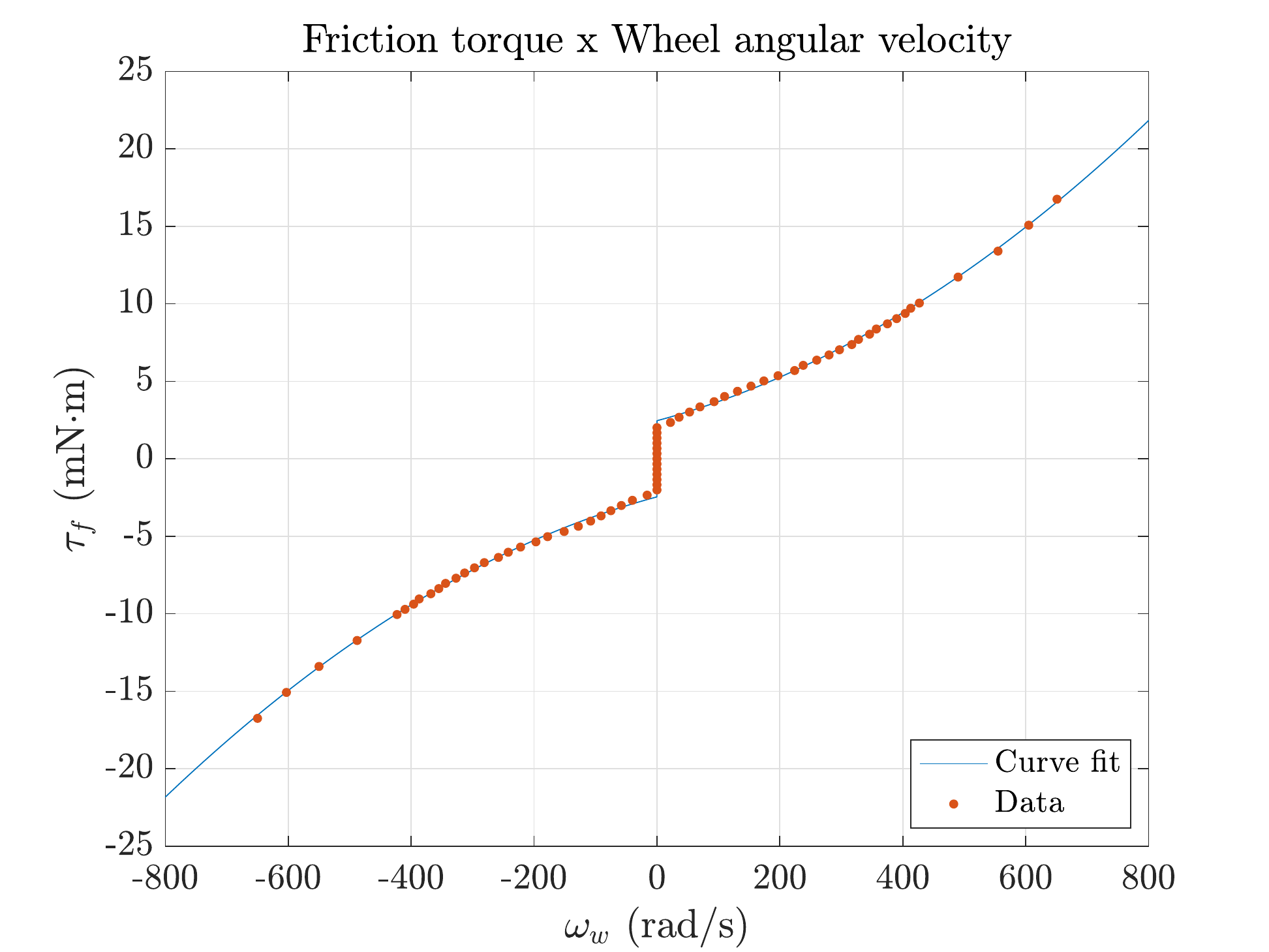}}
                \caption{Friction torque} 
                \label{fig/friction_torque_wheel_angular_velocity}
            \end{figure}

            Those parameters were determined experimentally with a torque controller by varying the torque reference, registering the equivalent steady-state velocity (where the input torque equals the friction torque) and then curve fitting the data (Fig. \ref{fig/friction_torque_wheel_angular_velocity}). The identified parameters are given in Tab. \ref{tab:friction_parameters}.
            \begin{table}[H]
                \caption{Friction torque parameters}
                \label{tab:friction_parameters}
                \centering
                \begin{tabular}{|cc|}
                    \hline
                    Parameter & Value \\
                    \hline
                    $\tau_c$ & $2.46\times10^{-3}$ N.m \\
                    $b_w$ & $1.06\times10^{-5}$ N.m.s.rad$^{-1}$ \\
                    $c_d$ & $1.70\times10^{-8}$ N.m.s$^2$.rad$^{-2}$ \\
                    \hline
                \end{tabular}
            \end{table}
            
        \subsection{Feedback linearization}
        
\begin{figure}[H]
                \centering
                \input{tikz/controller_feedback_linearization.tex}
                \caption{Cubli with feedback linearization}
                \label{fig:controller_feedback_linearization}
            \end{figure}
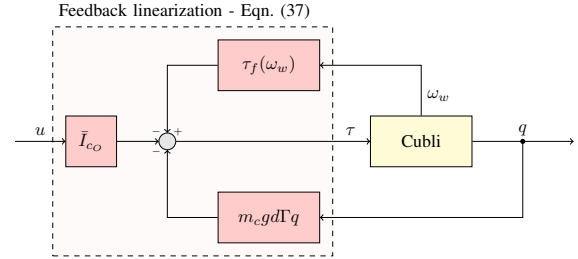

            Adopting a new input $u$ and making the input torque $\tau$ equal to:
            \begin{equation}
                \tau = - m_c g d \Gamma q + \tau_f ( \omega_w ) - \bar{I}_{c_O} u,
                \label{eqn:feedback_linearization}
            \end{equation}
            
            \noindent
            a feedback linearization law that cancels out the gravity torque and motor friction is obtained (Fig. \ref{fig:controller_feedback_linearization}).
            
            Substituting Eqn. (\ref{eqn:feedback_linearization}) into (\ref{eqn:dynamic_equations_complex_numbers}), reduces the system to:
            \begin{equation}
                \left\{
                \begin{array}{l}
                    \dot{q} = G(q)^T \omega_c \\
                    \dot{\omega}_c = u
                \end{array}
                \right.
                \label{eqn:dynamic_equations_linearized}
            \end{equation}
            
            Although the angular velocity differential equation is now linear, the unit complex number differential equation is still nonlinear.
    
        \subsection{State regulator}
        
            Let $q_r$ be an unit complex number reference and $q_e$ be an unit complex number error:
            \begin{equation}
                q_r =
                \begin{bmatrix}
                    q_{r_0} \\
                    q_{r_1}
                \end{bmatrix}
                , \qquad
                q_e =
                \begin{bmatrix}
                    q_{e_0} \\
                    q_{e_1}
                \end{bmatrix}
            \end{equation}
            
            Orientation error represents the rotation needed from current orientation to match orientation reference (Fig. \ref{fig/orientation_error}a):
            \begin{equation}
                \theta_r = \theta + \theta_e
            \end{equation}
            
            Because unit complex numbers always have unitary norm ($|q_r|=|q|=|q_e|=1$), in complex number notation, consecutive rotations can be represented as multiplications between respective complex numbers (Fig. \ref{fig/orientation_error}b), which means that:
            \begin{equation}
                q_r = q \circ q_e
                \label{eqn:qr}
            \end{equation}
            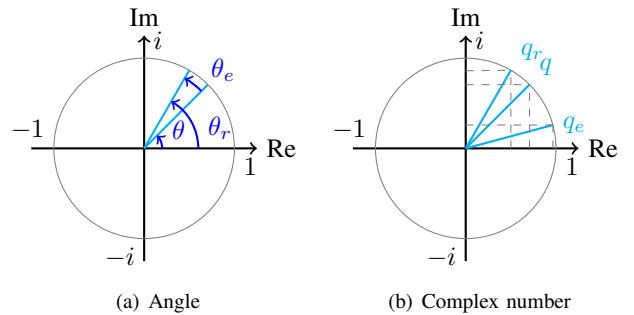
\begin{figure}[H]
                \centering
                \subfigure[][Angle]{\input{tikz/angle_error}}
                \subfigure[][Complex number]{\input{tikz/complex_number_error}} 
                \caption{Orientation error} 
                \label{fig/orientation_error}
            \end{figure}
            
            By left-multiplying both sides of Eqn. (\ref{eqn:qr}) with $\bar{q}$, it is possible to isolate unit complex number error $q_e$:
            \begin{align}
                %\bar{q} \circ q_r &= \cancel{\bar{q} \circ q} \circ q_e \nonumber \\
                q_e &= \bar{q} \circ q_r
                \label{eqn:qe}
            \end{align}
            
            When current orientation matches orientation reference, no additional rotation is needed and thus unit complex number error is $q_e = \begin{bmatrix} 1 & 0 \end{bmatrix}^T$. Because $q_e$ is not zero (and will never be, since an orientation complex number always have unitary norm), Eqn. (\ref{eqn:error_dynamics_0}) could not be used to guarantee asymptotically stable error dynamics:
            \begin{equation}
                    \ddot{q}_e + k_d \dot{q}_e + k_p q_e \neq \begin{bmatrix} 0 \\ 0 \end{bmatrix}
                    \label{eqn:error_dynamics_0}
            \end{equation}
            
            However, the imaginary part of the unit complex number error will be zero, which means that Eqn. (\ref{eqn:error_dynamics_2}) could be used instead:
            \begin{equation}
                    \ddot{q}_{e_1} + k_d \dot{q}_{e_1} + k_p q_{e_1} = 0
                    \label{eqn:error_dynamics_2}
            \end{equation}
        
            The first time derivative of $q_e$ can be calculated differentiating Eqn. (\ref{eqn:qe}) and making use of Eqn. (\ref{eqn/inverse_rotation_kinematic_equation}):
            \begin{align}
                \dot{q}_e &= \dfrac{d}{dt} \left( \bar{q} \circ q_r \right) \nonumber \\
                \dot{q}_e &= \dot{\bar{q}} \circ q_r + \bar{q} \circ \cancelto{0}{\dot{q}_r} \nonumber \\
                \dot{q}_e &= \dot{\bar{q}} \circ \left(  q \circ q_e \right) \nonumber \\
                \dot{q}_e &= - \begin{bmatrix} 0 \\ \omega_c \end{bmatrix} \circ q_e 
                %\quad \longrightarrow \quad
                \left\{
                \begin{array}{l}
                    \dot{q}_{e_0} = \omega_c q_{e_1} \\
                    \dot{q}_{e_1} = - \omega_c q_{e_0} 
                \end{array}
                \right.
                \label{eqn:qe_dot}
            \end{align}
            
            The second time derivative of $q_e$ can be calculated by differentiating Eqn. (\ref{eqn:qe_dot}):
            \begin{align}
                \ddot{q}_e &= \dfrac{d}{dt} \left( - \begin{bmatrix} 0 \\ \omega_c \end{bmatrix} \circ q_e  \right) \nonumber \\
                \ddot{q}_e &= - \begin{bmatrix} 0 \\ \dot{\omega}_c \end{bmatrix} \circ q_e - \begin{bmatrix} 0 \\ \omega_c \end{bmatrix} \circ \dot{q}_e \nonumber \\
                \ddot{q}_e &= - \begin{bmatrix} 0 \\ \dot{\omega}_c \end{bmatrix} \circ q_e - \begin{bmatrix} 0 \\ \omega_c \end{bmatrix} \circ \left( - \begin{bmatrix} 0 \\ \omega_c \end{bmatrix} \circ q_e \right) \nonumber \\
                \ddot{q}_e &= - \begin{bmatrix} 0 \\ \dot{\omega}_c \end{bmatrix} \circ q_e - \begin{bmatrix} \omega_c^2 \\ 0 \end{bmatrix} \circ q_e
                %\quad \longrightarrow \quad
                \left\{
                \begin{array}{l}
                    \ddot{q}_{e_0} = \dot{\omega}_c q_{e_1} - \omega_c^2 q_{e_0} \\
                    \ddot{q}_{e_1} = - \dot{\omega}_c q_{e_0} - \omega_c^2 q_{e_1} 
                \end{array}
                \right.
                \label{eqn:qe_ddot}
            \end{align}
            
            Substituting Eqn. (\ref{eqn:qe_dot}) and (\ref{eqn:qe_ddot}) into (\ref{eqn:error_dynamics_2}):
            \begin{align}
                \ddot{q}_{e_1} + k_d \dot{q}_{e_1} + k_p q_{e_1} &= 0 \nonumber \\
                \left( - \dot{\omega}_c q_{e_0} - \omega_c^2 q_{e_1} \right) + k_d \left( - \omega_c q_{e_0} \right) + k_p q_{e_1} &= 0 \nonumber \\
                \dot{\omega}_c + k_d \omega_c - \left( k_p - \omega_c^2 \right) \frac{q_{e_1}}{q_{e_0}} &= 0
            \end{align}
            
            Isolating $\omega_c$, yields the following control law:
            \begin{equation}
                u = \left( k_p - \omega_c^2 \right) \frac{q_{e_1}}{q_{e_0}} - k_d \omega_c
                \label{eqn:controller_law_linear}
            \end{equation}
            
            The term $\sigma_e = \frac{q_{e_1}}{q_{e_0}}$ is singular for $\pm 90^{\circ}$ rotations (since the term $q_{e_0} = \cos \theta_e$ appears in the denominator). Although this may appear to be a disadvantage, if it is necessary to go to a reference more than $90^{\circ}$ away, a trajectory control may be utilized. Moreover, for small rotations the term $\omega_c^2$ is close to zero and $q_{e_0}$ is close to one, which further simplifies the control law:
            \begin{equation}
                u \approx k_p q_{e_1} - k_d \omega_c
                \label{eqn:controller_law_linear_2}
            \end{equation}
            
            Also, for small rotations, the unit complex number error and angular velocity are approximate to:
            \begin{equation}
                q_e = 
                \begin{bmatrix}
                    \cos\theta_e \\
                    \sin\theta_e 
                \end{bmatrix}
                \approx
                \begin{bmatrix}
                    1 \\
                    \theta_{c_r} - \theta_c
                \end{bmatrix}
                , \qquad
                \omega_c \approx \dot{\theta}_c
                \label{eqn:qe_vec_euler_angles}
            \end{equation}
            
            Substituting Eqn. (\ref{eqn:qe_vec_euler_angles}) into (\ref{eqn:controller_law_linear_2}), yields a state regulator that is equal to the one commonly utilized with angles when dealing with small rotations:
            \begin{equation}
                u \approx k_p \left( \theta_{c_r} - \theta_c \right)  - k_d \dot{\theta}_c
            \end{equation}
            
            This means that, for small rotations, the derived nonlinear control law of Eqn. (\ref{eqn:controller_law_linear}) is equivalent to a linear one dynamically linearized at the reference.
        
        \subsection{Controller gains}
        
            Substituting Eqn. (\ref{eqn:controller_law_linear}) into (\ref{eqn:dynamic_equations_linearized}), and rewriting the first differential equation in terms of $\sigma_e$ instead of $q$, yields:
            \begin{equation}
                \left\{
                \begin{array}{l}
                    \dot{\sigma}_e = \left( 1 + \sigma_e^2 \right) \omega_c \\
                    \omega_c = \left( k_p - \omega_c^2 \right) \sigma_e - k_d \omega_c
                \end{array}
                \right.
            \end{equation}
            
            When the Cubli is in its equilibrium position $\sigma_e=\omega_c=0$, the closed-loop linearized dynamics are:
            \begin{equation}
                \begin{bmatrix}
                    \dot{\sigma}_e \\
                    \dot{\omega}_c
                \end{bmatrix}
                =
                \left[
                \begin{array}{c;{1pt/2pt}c}
                    0 & -1 \\
                    \hdashline[1pt/2pt]
                    k_p & -k_d
                \end{array}
                \right]
                \begin{bmatrix}
                    \sigma_e \\
                    \omega_c
                \end{bmatrix}
            \end{equation}
            
            Its characteristic polynomial is:
            \begin{equation}
                s^2 + k_d s + k_p = 0
                \label{eqn:characteristic_polynomial_1}
            \end{equation}
            
            Comparing Eqn. (\ref{eqn:characteristic_polynomial_1}) with the characteristic polynomial of a generic 2$^{\text{nd}}$ order system with two complex poles with damping ratio $\zeta$ and natural frequency $\omega_n$:
            \begin{equation}
                s^2 + 2 \zeta \omega_n s + \omega_n = 0
            \end{equation}
            
            \noindent
            yields the following values for the controller gains in terms of the desired closed-loop parameters $\zeta$ and $\omega_n$:
        
            \begin{equation}
                \left\{
                \begin{array}{l}
                    k_p = \omega_n^2 \\
                    k_d = 2 \zeta \omega_n
                \end{array}
                \right.
                \label{eqn:controller_gains_1}
            \end{equation}
            
    \section{ATTITUDE AND WHEEL CONTROLLER}
    
        Since the Cubli is influenced by the acceleration of the reaction wheel, it may happen that the reaction wheel velocity saturates for a while. Moreover, the attitude sensor may not be perfectly aligned with the Cubli, so what might appear to be an equilibrium position may actually not be, and the wheel will be always accelerating trying to keep the Cubli on that position. It is thus desirable to try to achieve the dual goals of stabilizing the Cubli and keep the wheel velocity small.
            \begin{strip}
            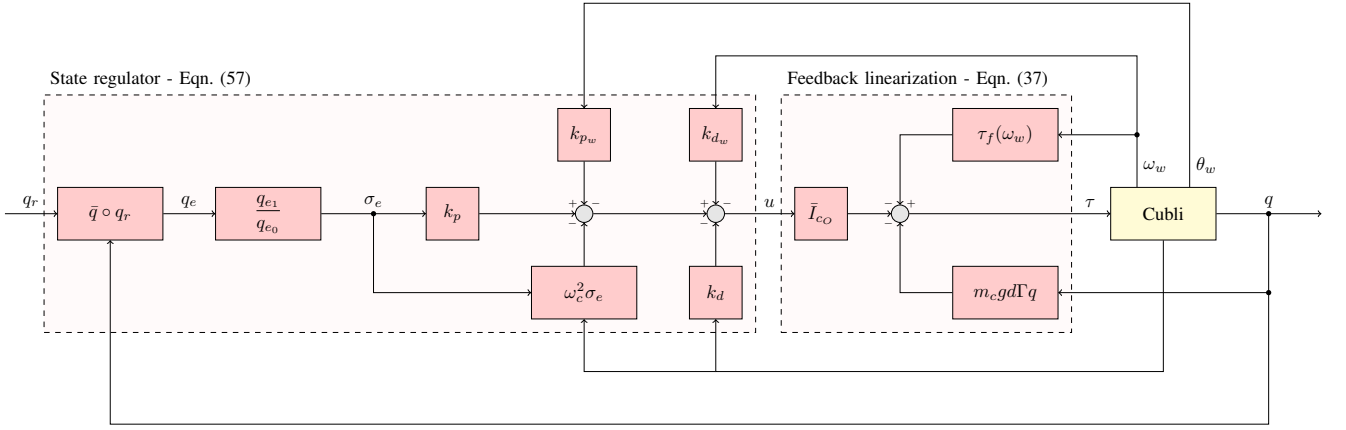
\begin{figure}[H]
                \centering
                \input{tikz/controller_state_regulator_with_feedback_linearization.tex}
                \caption{Cubli with state regulator and feedback linearization}
                \label{fig:controller_state_regulator_with_feedback_linearization_2}
            \end{figure}
            \end{strip}
        
        \subsection{State regulator}
        
            To achieve this, the control law of Eqn. (\ref{eqn:controller_law_linear}) may be slightly modified  by also having feedback from the reaction wheel angular displacement and velocity:
            \begin{equation}
                u = \left( k_p - \omega_c^2 \right) \sigma_e - k_d \omega_c - k_{p_w} \theta_w - k_{d_w} \omega_w
                \label{eqn:state_regulator_2}
            \end{equation}
            
            The full nonlinear control law (Fig. \ref{fig:controller_state_regulator_with_feedback_linearization_2}) is composed of the feedback linearization from Eqn. (\ref{eqn:feedback_linearization}) and the state regulator from Eqn. (\ref{eqn:state_regulator_2}).
        
        \subsection{Controller gains}
        
            Substituting Eqn. (\ref{eqn:state_regulator_2}) into (\ref{eqn:dynamic_equations_linearized}), and rewriting the first differential equation in terms of $\sigma_e$ instead of $q$, yields:
            \begin{equation}
                \left\{
                \begin{array}{l}
                    \dot{\sigma}_e = \left( 1 + \sigma_e^2 \right) \omega_c \\
                    \dot{\theta}_w = \omega_w \\
                    \omega_c = \left( k_p - \omega_c^2 \right) \sigma_e - k_d \omega_c - k_{p_w} \theta_w - k_{d_w} \omega_w \\
                    \dot{\omega}_w = \frac{m_c g d}{I_{w_G}} \sigma_e - k_p \frac{\bar{I}_{c_O}}{I_{w_G}} \sigma_e + k_d \frac{\bar{I}_{c_O}}{I_{w_G}} \omega_c \\ \hspace{2.5cm} + k_{p_w} \frac{\bar{I}_{c_O}}{I_{w_G}} \theta_w + k_{d_w} \frac{\bar{I}_{c_O}}{I_{w_G}} \omega_w
                \end{array}
                \right.
            \end{equation}
        
            When the Cubli is in its equilibrium position $\sigma_e=\omega_c=\theta_w=\omega_w=0$, the closed-loop linearized dynamics are:   
            \begin{equation}
                \resizebox{0.49\textwidth}{!}{$
                \begin{bmatrix}
                    \dot{\sigma}_e \\
                    \dot{\theta}_w \\
                    \dot{\omega}_c \\
                    \dot{\omega}_w
                \end{bmatrix}
                =
                \left[
                \begin{array}{c;{1pt/2pt}c;{1pt/2pt}c;{1pt/2pt}c}
                    0 & 0 & -1 & 0 \\
                    \hdashline[1pt/2pt]
                    0 & 0 & 0 & 1 \\
                    \hdashline[1pt/2pt]
                    k_p & -k_{p_w} & -k_d & -k_{d_w} \\
                    \hdashline[1pt/2pt]
                    \frac{m_c g d}{I_{w_G}} - k_p \frac{\bar{I}_{c_O}}{I_{w_G}} & k_{p_w} \frac{\bar{I}_{c_O}}{I_{w_G}} & k_d \frac{\bar{I}_{c_O}}{I_{w_G}} & k_{d_w} \frac{\bar{I}_{c_O}}{I_{w_G}} \\
                \end{array}
                \right]
                \begin{bmatrix}
                    \sigma_e \\
                    \theta_w \\
                    \omega_c \\
                    \omega_w
                \end{bmatrix}
                $}
            \end{equation}
            
            Its characteristic polynomial is:
            \begin{equation}
                s^4 + \left( k_d - \gamma k_{d_w} \right)s^3 + \left( k_p - \gamma k_{p_w} \right)s^2 + \delta k_{d_w} s + \delta k_{p_w} = 0
                \label{eqn:characteristic_polynomial_2}
            \end{equation}
            
            \noindent
            where:
            \begin{equation}
                \gamma = \frac{\bar{I}_{c_O}}{I_{w_G}}, \qquad \delta = \frac{m_cgd}{I_{w_G}}
            \end{equation}
            
            Comparing Eqn. (\ref{eqn:characteristic_polynomial_2}) with the characteristic polynomial of a generic 4$^{\text{th}}$ order system with two complex poles and two repeated real poles:
            \begin{align}
                \left( s^2 + 2\zeta\omega_ns + \omega_n^2 \right) {\left( s + \alpha\zeta\omega_n \right)}^2 &= 0 \nonumber \\
                s^4 + 2 \zeta \omega_n \left( 1+ \alpha \right) s^3 + \omega_n^2 \left( 1 + \alpha\zeta^2\left(4 + \alpha \right) \right) s^2 & \nonumber \\ + \left( 2\alpha\zeta\omega_n^3\left(1+\alpha\zeta^2\right) \right) s + \alpha^2\zeta^2\omega_n^4 &= 0
            \end{align}
            
            \noindent
            yields the following values for the controller gains in terms of the desired closed-loop parameters $\zeta$, $\omega_n$ and $\alpha$:
            \begin{equation}
                \left\{
                \begin{array}{l}
                    k_p = \omega_n^2 \left( 1 + \alpha \zeta^2 \left( 4 + \alpha \right)  \right) + \gamma \dfrac{\alpha^2\zeta^2\omega_n^4}{\delta} \\
                    k_d = 2 \zeta \omega_n \left( 1 + \alpha \right) + \gamma \dfrac{2\alpha\zeta\omega_n^3\left(1+\alpha\zeta^2\right)}{\delta} \\
                    k_{p_w} = \dfrac{\alpha^2\zeta^2\omega_n^4}{\delta} \\
                    k_{d_w} = \dfrac{2\alpha\zeta\omega_n^3\left(1+\alpha\zeta^2\right)}{\delta}
                \end{array}
                \right.
                \label{eqn:controller_gains_2}
            \end{equation}
            
            Note that, if $\alpha=0$, the controller gains $k_p$ and $k_d$ are equal to the ones derived in Eqn. (\ref{eqn:controller_gains_1}), while the controller gains $k_{p_w}$ and $k_{d_w}$ are equal to zero. By choosing a small enough value of $\alpha$, we guarantee that the reaction wheel dynamics would be slow enough to not interfere in the Cubli dynamics. In other words, the Cubli closed-loop poles will be sufficient faster than the reaction wheel closed-loop poles.
    
    \section{EXPERIMENTAL RESULTS}
    
        To validate the controller, experiments were realized with the Cubli prototype (Fig. \ref{fig/cubli}). Its electronics is composed of one STM32 NUCLEO-L432KC development board (80MHz ARM 32-bit Cortex M4), one SparkFun 9dof Sensor Stick inertial measurement unit (LSM9DS1), three Maxon EC 45 Flat brushless motors with a Maxon ESCON Module 50/5 dedicated motor controller each and one Turnigy Graphene Panther 1000mAh 6S LiPo battery. The microcontroller runs ARM Mbed OS open-source operating system, communicates with the IMU with I2C serial communication protocol and with the motor controllers with PWM and analog signals. A dedicated PCB was built to interface all these components. The mechanical parts were made in laser cut aluminum and 3D printed ABS.
    
\begin{figure}[H]
            \centering
            \includegraphics[width=0.475\textwidth]{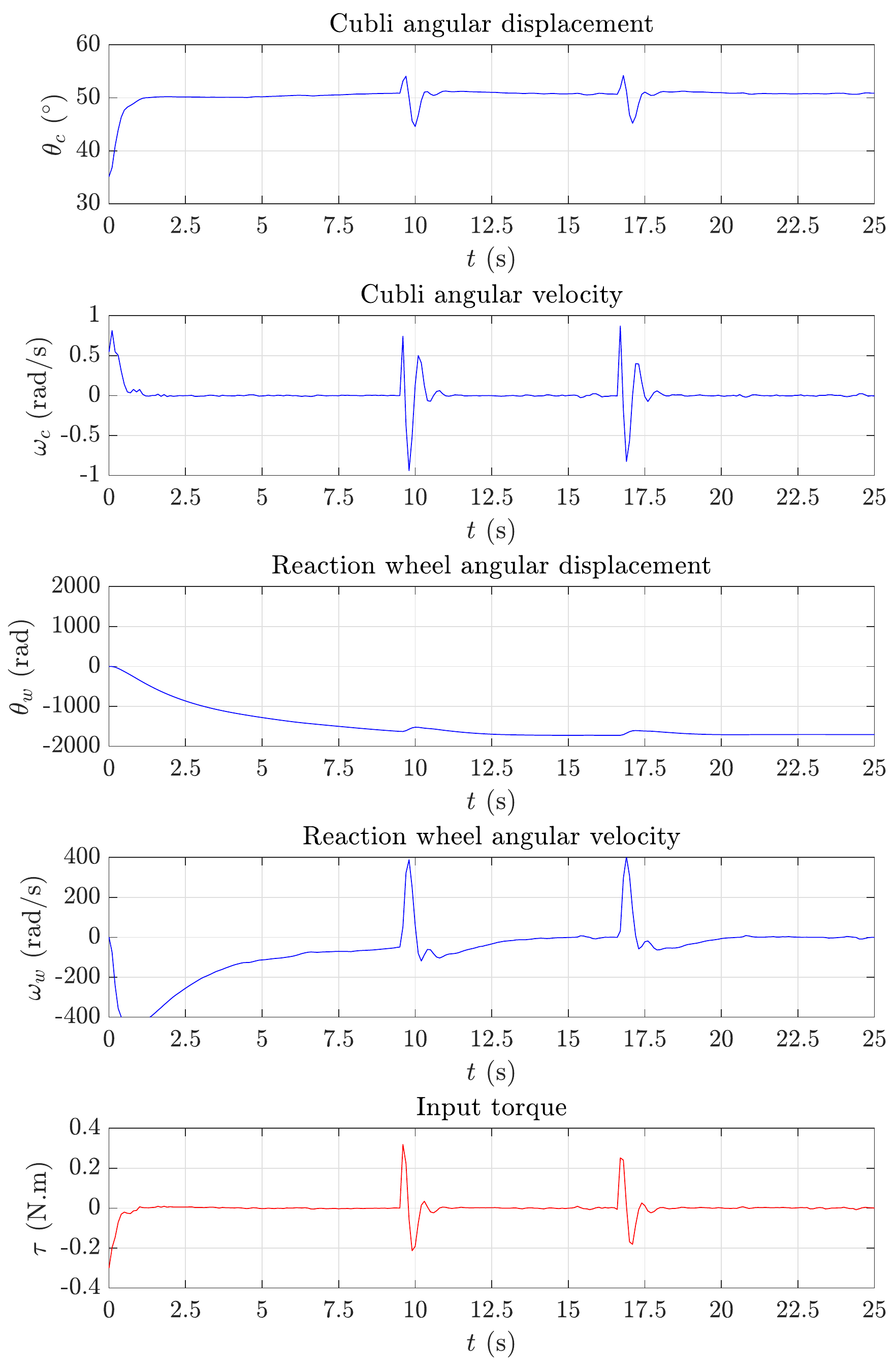}
            \caption{Experimental results}
            \label{fig:experimental_results}
        \end{figure}

        Experimental results were obtained (Fig. \ref{fig:experimental_results}), adopting $\zeta=\frac{\sqrt{2}}{2}$, $\omega_n = 1.5\omega_0$ and $\alpha = 0.1$ for the controller gains and setting the unit complex number reference to Cubli's unstable equilibrium position $q_r = q_u$, that is, $\theta_{c_r} = 45^{\circ}$.
        
        The Cubli was stabilized in less than 1 second, as can be seen for its angular velocity $\omega_c$ rapidly decaying to zero. The reaction wheel angular velocity $\omega_w$ also decayed to zero, but at a much slower rate of around 10 seconds. This makes total sense since $\alpha = 0.1$, which means the Cubli dynamics should be 10 times faster.
        
        Two disturbances were applied, one around 9 seconds and another around 16 seconds. It both cases the Cubli quickly re-stabilized itself without oscillating too much or saturating the actuators.
        
        Moreover, the Cubli did not stabilize at $45^{\circ}$ but at around $50^{\circ}$. This probably happened due to construction imperfections or sensor misalignment. However, because the reaction wheel states are also being feedbacked, the controller was able to find the real equilibrium position.
        
        A video of this and other experiments are available at \url{https://youtu.be/8krzqLFjemE}. 
    
    \section{CONCLUSION AND FUTURE WORK}
    
        By utilizing unit complex numbers instead of angles, a nonlinear control law was designed and implemented. This approach proved to be efficient given the experimental results of the Cubli balancing on its edge (1D). It has the advantage of not needing any trigonometric operations in the control algorithm, although the computation efficiency gains are not so significant due to recent advances in embedded microprocessors. The real advantage of this approach is to set the path of the nonlinear control law for the Cubli balancing on its vertex (3D), which uses unit ultra-complex numbers (quaternions) that, different from unit complex numbers, cannot be easily visualized and interpreted in our 3D world.
    
    %\nocite{*}
    \bibliographystyle{ieeetr}
    \bibliography{biblio}

\end{document}

%% file: packages.tex
\usepackage{url}

\usepackage{amsmath}
\usepackage{amsfonts}
\usepackage{arydshln}
\usepackage{cancel}

\usepackage{float}
\usepackage{subfigure}

\usepackage{cuted}

\usepackage{tikz}
\usepackage{tikz-3dplot}
\usetikzlibrary{fit}

\tikzstyle{plant_function} = [draw, rectangle, minimum height = 1cm, minimum width = 2cm, fill=yellow!20]
\tikzstyle{plant_gain} = [draw, rectangle, minimum height = 1cm, minimum width = 1cm, fill=yellow!20]
\tikzstyle{plant_container} = [draw, rectangle, dashed, inner sep=0.25cm, fill=yellow!2]
\tikzstyle{plant_int} = [draw, rectangle, minimum height = 1cm, minimum width = 1cm, fill=yellow!20]

\tikzstyle{controller_function} = [draw, rectangle, minimum height = 1cm, minimum width = 2cm, fill=red!20]
\tikzstyle{controller_gain} = [draw, rectangle, minimum height = 1cm, minimum width = 1cm, fill=red!20]
\tikzstyle{controller_container} = [draw, rectangle, dashed, inner sep=0.25cm, fill=red!2]

\tikzstyle{sum} = [draw, circle, minimum height = 0.25cm, minimum width = 0.25cm, fill=gray!20]

\pgfdeclarelayer{layer_container}
\pgfsetlayers{layer_container,main}

%% file: tikz/inverted_pendulum_cart_pole.tex
\tdplotsetmaincoords{0}{0}

\begin{tikzpicture}[tdplot_main_coords,scale=0.75]
    
    % Cart
    \draw[thick,fill=gray!50,rounded corners=2] (0.3,-0.2) -- (0.3,0.2) -- (-0.3,0.2) -- (-0.3,-0.2) -- cycle;
	\draw[thick,fill=gray!50] (-0.15,-0.3) circle [radius=0.1];
	\draw[thick,fill=gray!50] (0.15,-0.3) circle [radius=0.1];
	
	% Floor
    \draw[thick] (1.5,-0.4) -- (-1.5,-0.4);
    \fill[thick,fill=gray!50] (1.5,-0.6) -- (1.5,-0.4) -- (-1.5,-0.4) -- (-1.5,-0.6) -- cycle;
    
    % Pendulum
	\tdplotsetrotatedcoords{-30}{0}{0}	
    \draw[thick,tdplot_rotated_coords,fill=gray!50,rounded corners=2] (0.1,-0.1) -- (0.1,2.1) -- (-0.1,2.1) -- (-0.1,-0.1) -- cycle;
	\fill[] (0,0) circle [radius=0.05];
    
    % Verical axis
    \draw[thick,gray,dashed] (0,0) -- (0,2);
    
    % Inputs/outputs vectors
    \draw[thick,->,red] (-1.1,0) -- node[above]{$f$} (-0.3,0) ;
    \draw[thick,->,blue] (0,1.2) node[above right]{$\theta$} arc (90:65:1.2);

\end{tikzpicture}

%% file: tikz/inverted_pendulum_reaction_wheel.tex
\tdplotsetmaincoords{0}{0}

\begin{tikzpicture}[tdplot_main_coords,scale=0.75]
    
    % Floor
    \draw[thick,fill=gray!50,rounded corners=2] (0.3,-0.3) -- (0.3,0.2) -- (-0.3,0.2) -- (-0.3,-0.3) -- cycle;
    \fill[thick,fill=white] (1.5,-0.4) -- (1.5,-0.2) -- (-1.5,-0.2) -- (-1.5,-0.4) -- cycle;

    % Pendulum
	\tdplotsetrotatedcoords{-30}{0}{0}	
    \draw[tdplot_rotated_coords,thick,fill=gray!50,rounded corners=2] (0.1,-0.1) -- (0.1,2.1) -- (-0.1,2.1) -- (-0.1,-0.1) -- cycle;
	\fill[] (0,0) circle [radius=0.05];
	
	% Wheel
	\tdplotsetrotatedcoords{-30}{0}{0}	
	\draw[tdplot_rotated_coords,thick,fill=gray!50] (0,2,0) circle [radius=0.5];
	\fill[tdplot_rotated_coords] (0,2,0) circle [radius=0.05];
    
    % Verical axis
    \draw[thick,gray,dashed] (0,0) -- (0,2);
    
    % Inputs/outputs vectors
    \draw[tdplot_rotated_coords,thick,->,red] (0,2.6) node[right]{$\quad \tau$} arc (90:30:0.6);
    \draw[thick,->,blue] (0,1.2) node[above right]{$\theta$} arc (90:65:1.2);

\end{tikzpicture}

%% file: tikz/complex_number.tex
% \scalebox{0.75}{
\begin{tikzpicture}[scale=0.5]
    \draw[thick,->] (-2.5,0) -- (2.5,0) node[right]{Re};
    \draw[thick,->] (0,-2.5) -- (0,2.5) node[above]{Im};
    
    \draw[gray] (0,0) circle (2cm);
    \node[below right] at (2,0) {$1$};
    \node[above right] at (0,2) {$i$};
    \node[above left] at (-2,0) {$-1$};
    \node[below left] at (0,-2) {$-i$};
    
    \draw[thick,cyan] (0,0) -- (1.7321,1) node[above right]{$q$};
    
    \draw[dashed,gray] (1.7321,0) node[below]{$q_0$} -- (1.7321,1);
    \draw[dashed,gray] (0,1) node[left]{$q_1$} -- (1.7321,1);
    
    \draw[thick,->,blue] (1,0) node[above right]{$\theta$} arc (0:30:1) ;
\end{tikzpicture}
% }

%% file: tikz/cubli_diagram.tex
% Set angle of vision
\tdplotsetmaincoords{0}{0}

\begin{tikzpicture}[tdplot_main_coords,scale=0.75]

    % Set rotated reference frame
    \tdplotsetrotatedcoords{30}{0}{0}
	
	% Draw inertial reference frame
	\draw[->,thick,black] (0,0,0) -- (6,0,0) node[right]{$y$};
	\draw[->,thick,black] (0,0,0) -- (0,6,0) node[above]{$z$};
       
    % Draw cube and reaction wheel
    \draw[tdplot_rotated_coords,fill=gray!80,fill opacity=0.80] (0,0,0) -- (4,0,0) -- (4,4,0) -- (0,4,0) -- (0,0,0);
    \draw[tdplot_rotated_coords,fill=green,fill opacity=0.4] (2,2,0) circle(1.6);

	% Draw body reference frame
	%\draw[tdplot_rotated_coords,->,thick,red] (0,0,0) -- (6,0,0) node[anchor=west]{$y'$};
	%\draw[tdplot_rotated_coords,->,thick,red] (0,0,0) -- (0,6,0) node[anchor=south]{$z'$};

    % Draw angles
	\tdplotdrawarc[tdplot_rotated_coords,->,thick,red]{(2,2,0)}{1.8}{30}{60}{above}{$\tau$};

    % Draw angles
	\tdplotdrawarc[->,thick,blue]{(0,0,0)}{2}{0}{30}{right}{$\theta_c$};
	\tdplotdrawarc[tdplot_rotated_coords,->,thick,blue]{(2,2,0)}{0.8}{0}{30}{above right}{$\theta_w$};
    
    % Draw points O and G
    \fill (0,0,0) circle [radius=0.075] node[below left]{$O$};
    \fill[tdplot_rotated_coords] (2,2,0) circle [radius=0.075] node[above left]{$G$};
    
    % Draw auxiliary lines
	\draw[tdplot_rotated_coords,-,dashed] (0,0,0) -- node[above right]{$d$} (2,2,0);
	\draw[tdplot_rotated_coords,-,dashed] (2,2,0) -- (3.6,2,0);
	\draw[tdplot_rotated_coords,-,dashed] (2,2,0) -- (3.38,2.8,0);

    \draw[tdplot_rotated_coords,gray,->] (5.5,4.5,0) node[right,text width=1.5cm] 
        {Structure} to [out=180,in=60] (4.25,4.25,0);
        
    \draw[tdplot_rotated_coords,gray,->] (-1,2.25,0) node[left,text width=1.5cm] 
        {Reaction wheel} to [out=0,in=210] (0.25,2,0);

\end{tikzpicture}

%% file: tikz/plant_cubli_dynamics.tex
\scalebox{0.75}{
\begin{tikzpicture}[auto]

    % Blocks
	\coordinate (tau) at (0.5,0) {};
    \node[sum] (sum_1) at (2.5,0) {};
    \node[plant_gain] (I_c) at (4,0) {$\dfrac{1}{\bar{I}_{c_O}}$};
    \node[plant_int] (int_1) at (6,0) {$\dfrac{1}{s}$};
    \node[plant_int] (int_2) at (9,0) {$\dfrac{1}{s}$};
	\coordinate (theta_c) at (11.5,0) {};
    \node[sum] (sum_2) at (2.5,3) {};
    \node[plant_gain] (I_w) at (4,3) {$\dfrac{1}{I_{w_G}}$};
    \node[plant_int] (int_3) at (6,3) {$\dfrac{1}{s}$};
    \node[plant_int] (int_4) at (9,3) {$\dfrac{1}{s}$};
	\coordinate (theta_w) at (10.5,3) {};
    \node[plant_function] (tau_f) at (5,1.5) {$\tau_f(\omega_w)$};
    \node[plant_function] (gravity) at (5,-1.5) {$m_c g d \cos \theta_c $};
    
    % Coordinates
	\coordinate (tau_aux) at (1.5,0) {};
	\coordinate (theta_c_aux) at (10.5,0) {};
	\coordinate (omega_w_aux) at (7.5,3) {};
	\coordinate (tau_f_aux) at (2.5,1.5) {};

	% Nodes
	\fill (tau_aux) circle [radius=0.05];
	\fill (theta_c_aux) circle [radius=0.05];
	\fill (omega_w_aux) circle [radius=0.05];
	\fill (tau_f_aux) circle [radius=0.05];
	
	% Signals
    \node at (sum_1) [below left] {\tiny $-$};
    \node at (sum_1) [above left] {\tiny $-$};
    \node at (sum_1) [above right] {\tiny $+$};
    \node at (sum_2) [below left] {\tiny $-$};
    \node at (sum_2) [above left] {\tiny $+$};
    
    % Lines
    \draw[->] (tau) -- node[above, pos = 0.25]{$\tau$} (sum_1);
    \draw[->] (sum_1) -- (I_c);
    \draw[->] (I_c) -- node[above]{$\dot{\omega}_c$} (int_1);
    \draw[->] (int_1) -- node[above]{$\omega_c$} (int_2);
    \draw[->] (int_2) -- node[above,pos=0.75]{$\theta_c$} (theta_c);
    \draw[->] (theta_c_aux) |- (gravity);
    \draw[->] (gravity) -| (sum_1);
    \draw[->] (tau_aux) |- (sum_2);
    \draw[->] (sum_2) -- (I_w);
    \draw[->] (I_w) -- node[above]{$\dot{\omega}_w$} (int_3);
    \draw[->] (int_3) -- node[above]{$\omega_w$} (int_4);
    \draw[->] (int_4) -- node[above]{$\theta_w$} (theta_w);
    \draw[->] (omega_w_aux) |- (tau_f);
    \draw[->] (tau_f) -| (sum_2);
    \draw[->] (tau_f_aux) -- (sum_1);
    
    % Controller
    \begin{pgfonlayer}{layer_container}
		\node [plant_container, fit=(tau_aux) (I_w) (gravity) (theta_c_aux)] (sys) {};
        \node at (sys.north west) [above right] {Cubli - Eqn. (\ref{eqn:dynamic_equations})};
	\end{pgfonlayer}
    
\end{tikzpicture}
}

%% file: tikz/plant_cubli_dynamics_complex_numbers.tex
\scalebox{0.63}{
\begin{tikzpicture}[auto]

	\node at (1,4.5) {};

    % Blocks
	\coordinate (tau) at (0.5,0) {};
    \node[sum] (sum_1) at (2.5,0) {};
    \node[plant_gain] (I_c) at (4,0) {$\dfrac{1}{\bar{I}_{c_O}}$};
    \node[plant_int] (int_1) at (6,0) {$\dfrac{1}{s}$};
    \node[plant_function] (quat_mult) at (9,0) {$G(q)^T \omega_c$};
    \node[plant_int] (int_2) at (12,0) {$\dfrac{1}{s}$};
	\coordinate (q) at (14,0) {};
    \node[sum] (sum_2) at (2.5,3) {};
    \node[plant_gain] (I_w) at (4,3) {$\dfrac{1}{I_{w_G}}$};
    \node[plant_int] (int_3) at (6,3) {$\dfrac{1}{s}$};
    \node[plant_int] (int_4) at (9,3) {$\dfrac{1}{s}$};
	\coordinate (theta_w) at (10.5,3) {};
    \node[plant_function] (tau_f) at (5,1.5) {$\tau_f(\omega_w)$};
    \node[plant_function] (gravity) at (5,-1.5) {$m_c g d \Gamma q $};
    
    % Coordinates
	\coordinate (tau_aux) at (1.5,0) {};
	\coordinate (q_aux_1) at (13,0) {};
	\coordinate (q_aux_2) at (9,-1.5) {};
	\coordinate (omega_w_aux_1) at (7.5,3) {};
	\coordinate (tau_f_aux) at (2.5,1.5) {};

	% Nodes
	\fill (tau_aux) circle [radius=0.05];
	\fill (q_aux_1) circle [radius=0.05];
	\fill (q_aux_2) circle [radius=0.05];
	\fill (omega_w_aux_1) circle [radius=0.05];
	\fill (tau_f_aux) circle [radius=0.05];
	
	% Signals
    \node at (sum_1) [below left] {\tiny $-$};
    \node at (sum_1) [above left] {\tiny $-$};
    \node at (sum_1) [above right] {\tiny $+$};
    \node at (sum_2) [below left] {\tiny $-$};
    \node at (sum_2) [above left] {\tiny $+$};
    
    % Lines
    \draw[->] (tau) -- node[above, pos = 0.25]{$\tau$} (sum_1);
    \draw[->] (sum_1) -- (I_c);
    \draw[->] (I_c) -- node[above]{$\dot{\omega}_c$} (int_1);
    \draw[->] (int_1) -- node[above]{$\omega_c$} (quat_mult);
    \draw[->] (quat_mult) -- node[above]{$\dot{q}$} (int_2);
    \draw[->] (int_2) -- node[above,pos=0.75]{$q$} (q);
    \draw[->] (q_aux_1) |- (gravity);
    \draw[->] (gravity) -| (sum_1);
    \draw[->] (q_aux_2) -- (quat_mult);
    \draw[->] (tau_aux) |- (sum_2);
    \draw[->] (sum_2) -- (I_w);
    \draw[->] (I_w) -- node[above]{$\dot{\omega}_w$} (int_3);
    \draw[->] (int_3) -- node[above]{$\omega_w$} (int_4);
    \draw[->] (int_4) -- node[above]{$\theta_w$} (theta_w);
    \draw[->] (omega_w_aux_1) |- (tau_f);
    \draw[->] (tau_f) -| (sum_2);
    \draw[->] (tau_f_aux) -- (sum_1);
    
    % Controller
    \begin{pgfonlayer}{layer_container}
		\node [plant_container, fit=(tau_aux) (I_w) (gravity) (q_aux_1)] (sys) {};
        \node at (sys.north west) [above right] {Cubli - Eqn. (\ref{eqn:dynamic_equations_complex_numbers})};
	\end{pgfonlayer}
    
\end{tikzpicture}
}

%% file: tikz/controller_feedback_linearization.tex
\scalebox{0.675}{
\begin{tikzpicture}[auto]

    % Blocks
	\coordinate (u) at (0,0) {};
    \node[controller_gain] (I_c) at (1.5,0) {$\bar{I}_{c_O}$};
    \node[plant_function] (sys) at (8,0) {Cubli};
	\coordinate (q) at (11,0) {};
    \node[sum] (sum_1) at (3,0) {};
    \node[controller_function] (tau_f) at (5,1.5) {$\tau_f(\omega_w)$};
    \node[controller_function] (gravity) at (5,-1.5) {$m_cgd\Gamma q$};
    
    % Coordinates
	\coordinate (q_aux) at (10,0) {};
	
	% Nodes
	\fill (q_aux) circle [radius=0.05];
	
	% Signals
    \node at (sum_1) [below left] {\tiny $-$};
    \node at (sum_1) [above left] {\tiny $-$};
    \node at (sum_1) [above right] {\tiny $+$};
    
    % Lines
    \draw[->] (u) -- node[above]{$u$} (I_c);
    \draw[->] (I_c) -- (sum_1);
    \draw[->] (sum_1) -- node[above, pos = 0.9]{$\tau$} (sys);
    \draw[->] (sys) -- node[above]{$q$} (q);
    \draw[->] (sys) |- node[right, pos = 0.2]{$\omega_w$} (tau_f);
    \draw[->] (tau_f) -| (sum_1);
    \draw[->] (q_aux) |- (gravity);
    \draw[->] (gravity) -| (sum_1);
    
    % Controller
    \begin{pgfonlayer}{layer_container}
		\node [controller_container, fit=(I_c) (tau_f) (gravity)] (fbl) {};
        \node at (fbl.north west) [above right] {Feedback linearization - Eqn. (\ref{eqn:feedback_linearization})};
	\end{pgfonlayer}
    
\end{tikzpicture}
}

%% file: tikz/angle_error.tex
% \scalebox{0.65}{
\begin{tikzpicture}[scale=0.6]
    \draw[thick,->] (-2.5,0) -- (2.5,0) node[right]{Re};
    \draw[thick,->] (0,-2.5) -- (0,2.5) node[above]{Im};
    
    \draw[gray] (0,0) circle (2cm);
    \node[below right] at (2,0) {$1$};
    \node[above right] at (0,2) {$i$};
    \node[above left] at (-2,0) {$-1$};
    \node[below left] at (0,-2) {$-i$};
    
    \draw[thick,cyan] (0,0) -- (1.414,1.414);
    \draw[thick,cyan] (0,0) -- (1,1.732);

    \draw[thick,->,blue] (0.4,0) node[above right]{$\theta$} arc (0:45:0.4) ;
    \draw[thick,->,blue] (1.2,0) node[above right]{$\theta_r$} arc (0:60:1.2);
    \draw[thick,->,blue] (1.27,1.27) node[above right]{$\theta_e$} arc (45:60:1.8);

\end{tikzpicture}
% }

%% file: tikz/complex_number_error.tex
% \scalebox{0.65}{
\begin{tikzpicture}[scale=0.6]
    \draw[thick,->] (-2.5,0) -- (2.5,0) node[right]{Re};
    \draw[thick,->] (0,-2.5) -- (0,2.5) node[above]{Im};
    
    \draw[gray] (0,0) circle (2cm);
    \node[below right] at (2,0) {$1$};
    \node[above right] at (0,2) {$i$};
    \node[above left] at (-2,0) {$-1$};
    \node[below left] at (0,-2) {$-i$};
    
    \draw[thick,cyan] (0,0) -- (1.414,1.414) node[above right]{$q$};
    \draw[thick,cyan] (0,0) -- (1,1.732) node[above right]{$q_r$};
    \draw[thick,cyan] (0,0) -- (1.932,0.518) node[right]{$q_e$};
    
    \draw[dashed,gray] (1.414,0) -- (1.414,1.414);
    \draw[dashed,gray] (0,1.414) -- (1.414,1.414);
    \draw[dashed,gray] (1,0) -- (1,1.732);
    \draw[dashed,gray] (0,1.723) -- (1,1.732);
    \draw[dashed,gray] (1.932,0) -- (1.932,0.518);
    \draw[dashed,gray] (0,0.518) -- (1.932,0.518);
    
    %\draw[thick,->,blue] (0.5,0) node[above right]{$\theta$} arc (0:45:0.5) ;
    %\draw[thick,->,blue] (1,0) node[above right]{$\theta_r$} arc (0:60:1);
    %\draw[thick,->,blue] (1.5,0) node[above right]{$\theta_e$} arc (0:15:1.5);

\end{tikzpicture}
% }

%% file: tikz/controller_state_regulator_with_feedback_linearization.tex
\scalebox{0.7}{
\begin{tikzpicture}[auto]

	\node at (0,4.5) {};

    % Linear quadractic regulator
	\coordinate (q_r) at (0,0) {};
    \node[controller_function] (q_e) at (2,0) {$\bar{q} \circ q_r$};
    \node[controller_function] (q_e_vec) at (5,0) {$\dfrac{q_{e_1}}{q_{e_0}}$};
    \node[controller_gain] (k_p) at (8.5,0) {$k_p$};
    \node[sum] (sum_1) at (11,0) {};
    \node[sum] (sum_2) at (13.5,0) {};
    \node[controller_function] (omega_c_omega_c) at (11,-1.5) {$\omega_c^2\sigma_e$};
    \node[controller_gain] (k_d) at (13.5,-1.5) {$k_d$};
    \node[controller_gain] (k_p_w) at (11,1.5) {$k_{p_w}$};
    \node[controller_gain] (k_d_w) at (13.5,1.5) {$k_{d_w}$};
    
    % Feedback linearization
    \node[controller_gain] (I_c) at (15.5,0) {$\bar{I}_{c_O}$};
    \node[sum] (sum_3) at (17,0) {};
    \node[controller_function] (tau_f) at (19,1.5) {$\tau_f(\omega_w)$};
    \node[controller_function] (gravity) at (19,-1.5) {$m_cgd\Gamma q$};
    
    % Cubli
    \node[plant_function] (sys) at (22,0) {Cubli};
	\coordinate (q) at (25,0) {};
    
    % Auxiliary coordinates
	\coordinate (q_aux_1) at (24,0) {};
	\coordinate (q_aux_2) at (24,-1.5) {};
	\coordinate (q_aux_3) at (24,-4) {};
	\coordinate (omega_c_aux_1) at (13.5,-3) {};
	\coordinate (omega_w_aux_1) at (21.5,1.5) {};
	\coordinate (omega_w_aux_2) at (21.5,3) {};
	\coordinate (theta_w_aux_1) at (22.5,4) {};
	\coordinate (sigma_aux_1) at (7,0) {};
	
	% Nodes
	\fill (q_aux_1) circle [radius=0.05];
	\fill (q_aux_2) circle [radius=0.05];
	\fill (omega_w_aux_1) circle [radius=0.05];
	\fill (sigma_aux_1) circle [radius=0.05];
	
	% Signals
    \node at (sum_1) [below left] {\tiny $-$};
	\node at (sum_1) [above left] {\tiny $+$};
	\node at (sum_1) [above right] {\tiny $-$};
    \node at (sum_2) [below left] {\tiny $-$};
	\node at (sum_2) [above left] {\tiny $+$};
	\node at (sum_2) [above right] {\tiny $-$};
    \node at (sum_3) [below left] {\tiny $-$};
    \node at (sum_3) [above left] {\tiny $-$};
    \node at (sum_3) [above right] {\tiny $+$};
    
    % Lines
    \draw[->] (q_r) -- node[above]{$q_r$} (q_e);
    \draw[->] (q_e) -- node[above]{$q_e$} (q_e_vec);
    \draw[->] (q_e_vec) -- node[above]{$\sigma_e$} (k_p);
    \draw[->] (k_p) -- (sum_1);
    \draw[->] (sum_1) -- (sum_2);
    \draw[->] (sum_2) -- node[above, pos = 0.64]{$u$} (I_c);
    \draw[->] (I_c) -- (sum_3);
    \draw[->] (sum_3) -- node[above, pos = 0.9]{$\tau$} (sys);
    \draw[->] (sys) -- node[above]{$q$} (q);
    \draw[->] (sys.north -| omega_w_aux_1) |- node[right, pos = 0.2]{$\omega_w$} (tau_f);
    \draw[->] (tau_f) -| (sum_3);
    \draw[->] (q_aux_1) |- (gravity); 
    \draw[->] (gravity) -| (sum_3);
    \draw[->] (sys) |- (omega_c_aux_1) -- (k_d);
    \draw[->] (k_d) -- (sum_2);
    \draw[->] (omega_c_aux_1) -| (omega_c_omega_c);
    \draw[->] (sigma_aux_1) |- (omega_c_omega_c);
    \draw[->] (omega_c_omega_c) -- (sum_1);
    \draw[->] (q_aux_2) -- (q_aux_3) -| (q_e); 
    \draw[->] (omega_w_aux_1) -- (omega_w_aux_2) -| (k_d_w);
    \draw[->] (k_d_w) -- (sum_2);
    \draw[->] (sys.north -| theta_w_aux_1) -- node[right, pos = 0.12]{$\theta_w$} (theta_w_aux_1) -| (k_p_w);
    \draw[->] (k_p_w) -- (sum_1);
    
    % State regulator
    \begin{pgfonlayer}{layer_container}
		\node [controller_container, fit=(q_e) (k_d_w) (k_d)] (lqr) {};
        \node at (lqr.north west) [above right] {State regulator - Eqn. (\ref{eqn:state_regulator_2})};
	\end{pgfonlayer}
    
    % Feedback linearization
    \begin{pgfonlayer}{layer_container}
		\node [controller_container, fit=(I_c) (tau_f) (gravity)] (fbl) {};
        \node at (fbl.north west) [above right] {Feedback linearization - Eqn. (\ref{eqn:feedback_linearization})};
	\end{pgfonlayer}
    
\end{tikzpicture}
}